\documentclass[]{interact}

\usepackage{epstopdf}
\usepackage[caption=false]{subfig}
\usepackage{float}

\usepackage{color}

\usepackage[numbers,sort&compress]{natbib}
\bibpunct[, ]{[}{]}{,}{n}{,}{,}
\renewcommand\bibfont{\fontsize{10}{12}\selectfont}

\theoremstyle{plain}

\theoremstyle{definition}

\theoremstyle{remark}

\begin{document}


\title{Heuristic Algorithm for Univariate Stratification Problem}

\author{
\name{J.A.M. Brito \textsuperscript{a}\thanks{CONTACT J.A.M.Brito. Email: jambrito@gmail.com}, G.S. Semaan\textsuperscript{b}, L. de Lima\textsuperscript{c} and A.C.Fadel \textsuperscript{d}}
\affil{\textsuperscript{a}Escola Nacional de Ciências Estatísticas – ENCE/IBGE, Rio de Janeiro, Brazil; \textsuperscript{b}Instituto do Noroeste Fluminense de Educação Superior (INFES) da Universidade Federal Fluminense (UFF), Rio de Janeiro, Brazil;
\textsuperscript{c} Universidade Federal do Paraná – UFPR, Paraná, Brazil;
\textsuperscript{d} Instituto Brasileiro de Geografia e Estatística – IBGE, Rio de Janeiro, Brazil}
}

\maketitle

\begin{abstract}
In sampling theory, stratification corresponds to a technique used in surveys, which allows segmenting a population into homogeneous subpopulations (strata) to produce statistics with a higher level of precision. In particular, this article proposes a heuristic to solve the univariate stratification problem - widely studied in the literature. One of its versions sets the number of strata and the precision level and seeks to determine the limits that define such strata to minimize the sample size allocated to the strata. A heuristic-based on a stochastic optimization method and an exact optimization method was developed to achieve this goal. The performance of this heuristic was evaluated through computational experiments, considering its application in various populations used in other works in the literature, based on 20 scenarios that combine different numbers of strata and levels of precision. From the analysis of the obtained results, it is possible to verify that the heuristic had a performance superior to four algorithms in the literature in more than 94\% of the cases, particularly concerning the known algorithms of Kozak and Lavallé-Hidiroglou.
\end{abstract}

\begin{keywords}
Stratification; Minimum Sample; Allocation; Optimization; Algorithms  
\end{keywords}

\section{Introduction}

Official statistical institutes make available, with great frequency, a wide range of data and statistics of great relevance to governments, with regard, for example, to the planning and implementation of their public policies, as well as to the citizen willing to accompany them. For the most part, these data are produced from surveys based on probabilistic sampling \cite{Lohr19}.

According to \cite{Lohr19} and \cite{Cochran77}, sampling techniques allow the collection and production of statistics of interest in less time, with lower cost and greater accuracy. However, when are considered administrative issues, and there is a requirement of the desired minimum precision about the statistics produced, it is necessary to incorporate stratification in the planning/design phase of the sample in search of the least possible variability. Stratified sampling is well used together with other techniques such as simple random sampling and cluster sampling.

When solving the stratification problem (univariate in this article), the $N$ observations of an $X$ variable (stratification variable), known for the entire population, are distributed among $L$ population strata by determining $L-1$ cutoff points, called strata boundaries. To determine such cutoff points, one of the following objectives should be considered: (i) setting a sample of size $n$, minimizing a variance expression (or the coefficient of variation) associated with variable $X$, or (ii) setting a level of precision (the target coefficient of variation) associated with variable $X$, minimize the total sample size $n$ that will be allocated to $L$ strata.

Determining the cutoff points considering one of these objectives (versions) is equi\-valent to solving a nonlinear integer optimization problem that is difficult to solve computationally. Therefore, in recent decades, several methods have been proposed, basically classified as approximation or optimization, and most of them has considered objective (i), \cite{Semaan20}, \cite{Hidiroglou2018}.

This article considers a hybrid approach that works to identify the cutoff points of the strata through a stochastic optimization method and then, through an exact method proposed by \cite{Brito2015}, perform the optimal allocation of the sample to the strata. More specifically, it is proposed a heuristic algorithm based on the Biased Random Key Genetic Algorithms (BRKGA)  meta-heuristic, \cite{Goncalves2011}, for the univariate stratification problem for objective (ii).

The article is organized as follows: Section \ref{USP} presents the univariate stratification problem, considering these two possible objectives, as well as a literature review regarding the main methods applied to its resolution. Section \ref{PDH} brings a discretization proposal for this problem and the proposed heuristic. Finally, Section \ref{NE} presents the results and analyzes considering a set of 31 populations and 20 different scenarios in relation to the number of strata and levels of precision.

\section{Univariate Stratification Problem} \label{USP}

Consider a population $P$ with $N$ units defined from a set $U = \{1,2, \dots ,N\}$ and the stratification variable $X$. Then, to obtain a good estimate $t_x$ for the variable $X$ and a set of $Y$ variables investigated in the survey, the population of total size $N$ is divided into $L$ subpopulations, with $N_h$ units (where ${h = 1, \dots, L}$ and ${N_1 + \cdots + N_L = N}$) called population strata \cite{Lohr19} and denoted by $U_1,U_2,\dots,U_L$. 

To carry out this division, the population vector $X_U=(x_1,x_2,...,x_N)$ with $N$ values of $X$ associated to the population units is defined such that $x_1\le x_2\le \ldots \le x_N$. Population strata are determined using $(L-1)$ cutoff points $b_1<\ldots<b_h<\ldots<b_{L-1}$ such that  $b_h\in[x_1,\ x_N]$  for each $h=1,\cdots,L-1$, and according to (\ref{eq:n4})-(\ref{eq:n6}):
\begin{equation}
U_1=\{i \in \ U:x_i\le b_1\}
\label{eq:n4}
\end{equation}
\begin{equation}
U_h=\{i  \in  U:b_{h-1} <x_i \le b_h\}, h=2,3,\dots,L-1
\label{eq:n5}
\end{equation}
\begin{equation}
U_L=\{i \in U: b_{L-1}<x_i\}
\label{eq:n6}
\end{equation}

From the determination of the $U_h$ (${h = 1, \dots, L}$) strata it is immediately observed that ${|U_1| + \cdots + |U_L| = N}$ and the values of $N_h$ and $S_{xh}^2$, corresponding respectively to population size and population variance of $X$ in stratum $h$, are obtained. In a later step, considering the fixed sample size ($n$) or a level of precision (target coefficient of variation - $cv_t$) associated with the estimator of the variable $X$, the allocation \cite{Lohr19,Cochran77} of a sample of size $n_h$ $(h=1,\dots ,L)$ is done.

When solving the univariate stratification problem, one must determine the cutoff points $b_1,b_2,...,b_{L-1}$ and the values of $n_h$ (where ${n_1 + \cdots + n_L = n}$ and ${2 \le n_h \le N_h} ~\forall~ {h = 1, \dots, L}$), which correspond to the sample allocation, that is, the size of the sample to be independently selected in each stratum, to obtain the minimum variance (or the coefficient of variation), fixed sample size $n$; or obtain the minimum sample size for a fixed level of precision, which corresponds to solving one of the following two optimization problems:

\vspace{0.3cm}

\noindent \textbf{Problem I}

\begin{equation}
Minimize \; V(t_x)=\sum_{h=1}^{L}N_h^2\frac{S_{xh}^2}{n_h}(1-\frac{n_h}{N_h})
\label{eq:n7}
\end{equation}
\hspace{3.7cm} \textit{Subject to:}
\begin{equation}
n_1 + \cdots +n_L = n
\label{eq:n8}
\end{equation}
\begin{equation}
2 \le n_h \le N_h , h=1,\dots,L
\label{eq:n9}
\end{equation}
\begin{equation}
n_h \in Z_+  , h=1,\dots,L
\label{eq:n10}
\end{equation}

\vspace{0.3cm}

\noindent \textbf{Problem II}

\vspace{0.3cm}

\begin{equation}
Minimize \ n=\sum_{h=1}^{L}n_h
\label{eq:n11}
\end{equation}
\hspace{5.3cm} \textit{Subject to:}
\begin{equation}
cv(t_x) \le cv_t
\label{eq:n12}
\end{equation}
\begin{equation}
2 \le n_h \le N_h , h=1,\dots,L
\label{eq:n13}
\end{equation}
\begin{equation}
n_h \in Z_+  , h=1,\dots,L
\label{eq:n14}
\end{equation}

It is necessary to determine the cutoff points and the sample sizes $n_h$ (decision variables) that minimize the respective objective functions and meet the constraints of these problems to obtain the optimum concerning Problem (I) or (II). The nonlinearity of Problem (I) is present in the objective function in \eqref{eq:n7}, and the nonlinearity of Problem (II) is associated with the restriction of the precision  level in \eqref{eq:n12}.

According to several approaches in the literature proposed for solving (I) and (II), after determining the strata, it is common to apply the Neyman allocation (\ref{eq:n15}) \cite{Lohr19} to determine the sample sizes of the Problem (I). The Equation (16) used in Problem (II) is obtained from the substitution of (15) in the term on the left side of equation (12) and algebraic manipulations.

\begin{equation}
n_h=n \frac{S_{xh}N_h}{\sum_{g=1}^{L}{S_{xg}N_g}}, h=1,\dots,L
\label{eq:n15}
\end{equation}
\begin{equation}
n=\frac{{(\sum_{h=1}^{L}{N_hS_{xh}})}^2}{{cv}_t^2 T_x^2+\sum_{h=1}^{L}{N_hS_{xh}^2}}
\label{eq:n16}
\end{equation}

When using this allocation or other allocation methods in the literature \cite{Cochran77,Bankier1988}, non-integer sample sizes $n_h$, in general, are obtained. It is necessary to apply a rounding procedure to ensure that the integrality constraint is satisfied. The issue of doing so, is that such rounding can produce sample sizes that do not meet constraints (8) and (9) or (12) and (13), which are mandatory.

In order to address this issue, the heuristic proposed in Section \ref{PDH} for solving the Problem (II) considers, in the phase of allocation of sample sizes to strata, the application of the exact method proposed by [24]. This method guarantees the overall optimal concerning sample allocation. More specifically, values  $n_h \in Z_+ (h=1,...,L)$  such that the objective function of (II) has a minimum value and are fulfilled the constraints (12), (13), and (14).

\subsection{Literature Review }\label{class}

Due to the relevance of the univariate stratification problem and its computational complexity, several algorithms are proposed in the literature, classified as approximate or optimization, according to \cite{Hidiroglou2018}. Approximate algorithms are simple to implement, require few computational operations, and are not focused on searching for optimal cutoff points. These points, together sample size, allow the minimization of the obje\-ctive function of Problem (I) or (II), which implies low stratification efficiency.

On the other hand, algorithms based on optimization often make possible to determine solutions of superior quality if compared to those obtained from approximated methods, although it demands intensive computation. In \cite{Danish2017}, the authors present a review of several methods from 1950 to the present. Next, we present other references associated with the main stratification methods.

Regarding the class of algorithms that deal with  Problem (I), in \cite{Dalenius1959}, there is an algorithm based on the application of the Dalenius-Hodges rule. In \cite{Hedlin2000}, considering an extension of Ekman's rule \cite{Ekman1959}, the strata are determined so that the variance of the stratification variable is minimal, with pre-fixed $n$ and $L$ and application of the Neyman allocation. In \cite{Gunning2004}, a simple and easy-to-implement algorithm is proposed, which uses the general term formula of a geometric progression to determine the strata limits for asymmetric populations.

In \cite{Unnithan1978}, an algorithm based on the modified Newton method is proposed, which determines the limits of the strata and produces an optimal location about the variance expression, based on the Neyman allocation and the choice of a good starting point. In \cite{Kozak2006}, the authors evaluate the algorithms proposed in \cite{Kozak2004, Gunning2004}, and \cite{Lavallee1988}. Finally, in \cite{Khan2008}, under the hypothesis that $X$ has a normal or triangular distribution and that the sampling is done with replacement, an algorithm is proposed that determines the cutoff points of the strata, using the Dynamic Programming technique. This algorithm can be applied to minimize a variance expression, considering the use of uniform, proportional, or Neyman allocation.

In \cite{Brito2010}, an exact algorithm based on concepts of graph theory and the minimization of the expression of variance and application of proportional allocation is proposed. In \cite{Kozak2014}, a comparative study between the algorithm proposed in \cite{Kozak2004} and \cite{Kozak2006} and the algorithm proposed in \cite{Keskinturk2007} is presented. In \cite{Hidiroglou2018}, the authors compare optimization methods and approximate methods, considering the stratification of asymmetric populations. Finally, in \cite{Khan2015a} and \cite{Khan2015b}, algorithms based on the Dynamic Programming technique are proposed and simultaneously perform the stratification and allocation of the sample.

In \cite{Reddy2019}, under the hypothesis that the stratification variable has a Weibull distribution, the stratification problem is solved using the Dynamic Programming technique and Neyman allocation. In \cite{Keskinturk2007}, a genetic algorithm that considers the minimization of variance is proposed, based on four possible types of sample allocation. In \cite{Danish2018a} and \cite{Danish2018b}, an approach based on dynamic programming for the two-variable stratification problem uses proportional and Neyman allocations. Finally, in \cite{Danish2019}, the bivariate stratification problem is also solved via dynamic programming, evaluating different probability distribution functions.

In \cite{Brito2010b}, an algorithm is proposed based on the ILS (Iterated Local Search) method and Neyman's allocation. In \cite{Brito2019} and \cite{Brito2017}, are proposed algorithms based, respectively, on the  BRKGA (Biased Random Key Genetic Algorithm)  and GRASP (Greedy Randomized Adaptive Search Procedure) methods, using the exact method proposed in \cite{Brito2015} in the allocation phase.

About the set of works that address the Problem (II), in \cite{Lavallee1988}, an algorithm for stra\-tification of asymmetric populations was proposed that uses the power allocation \cite{Bankier1988} to determine the sample size. In \cite{Rivest2002}, is proposed a generalization of the algorithm presented in \cite{Lavallee1988}, which incorporates two models that account for the discrepancy between the stratification variable and a research study variable. In \cite{Kozak2004}, an algorithm called Random Search is proposed, which uses optimization concepts to determine the cutoff points of the strata and the Neyman allocation. In \cite{BR2009} several methods based on Kozak's \cite{Kozak2004} and Sethi's \cite{Sethi1963} algorithms are compared, concluding that the former, with initial strata of equal size, is more reliable and efficient than other approaches taken into account, and \cite{BR2011} introduces the R package \textit{stratification} that implements methods used in \cite{BR2009}, in addition to allowing to consider stratum-specific anticipated non-response and a model for the relationship between stratification and survey variables.

In \cite{Lisic2018}, an algorithm is proposed that solves the multivariate stratification problem and uses a penalized objective function, which is optimized by applying the Simulated Annealing method. In \cite{Brito2021}, the authors propose a brute force algorithm, limited to small populations ($N$ value), and an algorithm based on the VNS (Variable Neighborhood Search) method, where, in the sample allocation step, both do use of the exact method proposed by \cite{Brito2015}. As a result of this method, the authors developed the R package \emph{stratvns}. In \cite{Ballin2013}, a genetic algorithm is proposed to solve the multivariate stratification problem, with possible  application to the univariate case, which originated the R package \textit{SamplingStrata}. In \cite{RK2020}, the \textit{stratifyR} package is designed by implementing a dynamic programming algorithm proposed by \cite{KKA2002}, \cite{Khan2015a} and \cite{Khan2008} to solve the optimal strata boundaries and the optimum sample sizes for the univariate problem. The \emph{stratification} package \cite{BR2011} implements a generalized Lavallee-Hidiroglou method \cite{Lavallee1988} of strata construction. 

It is worth mentioning that we have used the R packages: \textit{stratification}, \cite{BR2011},  \textit{stratvns}, \cite{Brito2021}, and \textit{SamplingStrata}, \cite{Ballin2013} in the numerical experiments section, and their results compared to the BRKGA metaheuristic. 

\section{Problem Discretization and Heuristc} \label{PDH}

\subsection{Discretization} \label{discre}

Before detailing the proposed heuristic, the discretization also considered in \cite{Brito2017} and \cite{Brito2010} will be presented and used as a basis for representing the solutions produced by the heuristic.

From the definition of the stratification problem and the Equations (\ref{eq:n4}), (\ref{eq:n5}), and (\ref{eq:n6}) presented in Section \ref{USP}, it is possible, initially, to take the values of the vector $X_U$ and gather it into a set $B$ considering only the distinct values.

Then, from $B$, it is possible to determine a possible partition of $X_U$ by the $L$ population strata, more specifically, how many and which $X_U$ values are in each stratrum. Assumed the basic and necessary assumption (associated with constraints (\ref{eq:n9}) and (\ref{eq:n13})) that $|B| \ge 2L$, guaranteeing at least one solution to the problem, this co\-rresponds to determining a non-negative integer solution that satisfies constraints (\ref{eq:n17}) and (\ref{eq:n18}):
\begin{equation}
w_1 + \cdots + w_h + \cdots + w_L=|B|
\label{eq:n17}
\end{equation}
\begin{equation}
w_h \ge 2,  \ h =1,\dots,L
\label{eq:n18}
\end{equation}

In these constraints, each $w_h$ corresponds to the number of observations of $B$ that are in the $h$-th stratum. Constraint (\ref{eq:n17}) guarantees that the sum of the total observations ($N_h$) in each of the strata corresponds to $N$, and constraint (\ref{eq:n18}) implies that each stratum has at least two distinct observations.

From each solution $(w_1, \dots, w_L)$ that satisfies (\ref{eq:n17}) and (\ref{eq:n18}), the corresponding values of $N_h$ and $S_{xh}^2$ are obtained. Then, using equations (\ref{eq:n16}) and (\ref{eq:n15}),  the values of $n_h$ (already rounded) are determined. In the next step, the values of $n_h$,  $N_h$, and $S_{xh}^2$ are used to calculate the objective function of the Problem (II) and to assess whether constraints (\ref{eq:n12}) and (\ref{eq:n13}) are attended.

For example, assuming  $L=3$, $X_U=\{1,2,3,3,4,5,7,8,8,9,10,12,12,15\}$ and using (\ref{eq:n17}) and (\ref{eq:n18}), are obtained  $B=\{1,2,3,4,5,7,8,9,10,12,15\}$ and Equations (\ref{eq:n19}) and (\ref{eq:n20}):
\begin{equation}
w_1+w_2+w_3=11 
\label{eq:n19}
\end{equation}
\begin{equation}
w_1\ge , w_2\ge  , w_3 \ge 2
\label{eq:n20}
\end{equation}

Analyzing (\ref{eq:n19}) and (\ref{eq:n20}), we have that $w=(4,5,2)$ is a possible solution. Since $w_1=4$, this implies that stratum 1 has the values 1, 2, 3, and 4, including the repeated values. Consequently, by the correspondence between $B$ and $X_U$, one arrives at $N_1=5$ and, similarly, $w_2=5$ implies $N_2=6$ and $w_3=2$ implies $N_3=3$. This correspondence also makes it possible to calculate the $S_{xh}^2$ variances.

From this discretization, solving the univariate stratification problem to obtain the optimum consists of determining, among all the possible solutions that satisfy (\ref{eq:n17}) and (\ref{eq:n18}), the one that simultaneously produces the cutoff points and the sample size that minimizes the objective function of the Problem (II). According to \cite{Brito2021}, the total of integer solutions of (\ref{eq:n17}) and (\ref{eq:n18}) is defined by: 
\begin{equation}
TS=\frac{(|B|-L-1)!}{(|B|-2L)!\ (L-1)!}
\label{eq:n21}
\end{equation}

Also, in \cite{Brito2021}, the authors proposed a brute force algorithm (AFB) that enumerates all solutions $w$ that satisfies, simultaneously (\ref{eq:n17}) and (\ref{eq:n18}). For each solution $w$, the algorithm determines the associated values of $N_h$ and $S_{xh}^2$ $(h=1,\dots , L)$ and, in a later step, applies the method proposed in \cite{Brito2015} to obtain the sample allocation. The solution $w$ corresponding to the smallest sample size is the optimal solution to the Problem (II). However, according to \cite{Brito2021}, even considering all the computational apparatus currently available, the AFB has its application restricted to cases where $TS \le 10^7$, a limit that is exceeded for small values, such as $N = 160$ and $L = 3$, for example. It is the main focus of the Computational Intractability theme, as exposed in \cite{Semaan20}.

Due to the complexity of the stratification problem and the impossibility of applying the AFB to stratify populations with a more significant number of observations, this section presents a heuristic for the stratification problem corresponding to the optimization model defined in Problem (II), which had as based on this discretization and the application of a stochastic optimization method known as BRKGA.

\subsection{BRKGA} \label{brkga}

The biased random-key genetic algorithm (BRKKA) is a stochastic optimization method, also known as metaheuristics \cite{Marti2018} and \cite{Goncalves2011}, that has been applied in several optimization problems, for example, in \cite{Fadel2021,Festa2013} and which has many similarities with genetic algorithms (GAs), which are a particular class of evolutionary algorithms \cite{Marti2018}.

In each generation (iteration) of the BRKGA, $p$ solutions (vectors) associated with the optimization problem in question and corresponding to the BRKGA population are produced, evaluated (via objective function), and updated. Furthermore, to produce new solutions during BRKGA generations, ensuring diversity and quality, are applied procedures called selection, crossover, and mutation operators in each generation of this method.

To apply the selection and crossover operators, it is necessary to calculate the value of the objective function of the problem in question at each generation for each $p$ solution (current population), ordering them increasingly (minimization problem) according to the objective function value. Then, considering such ordering, the best $p_e$ solutions are allocated to a set of elite solutions denoted by $S_e$. Finally, the other solutions $(p - p_e)$ are allocated to a set of non-elite solutions $S_{ne}$.

The selection operator is applied to ensure that the best solutions (i.e., solutions with the highest objective function value) will be copied in the population for all generations. More specifically, are copied to the population  $p_e$ vectors of $S_e$, considered in the next generation. Finally, the other solutions of this new population are obtained from the application of the crossover and mutation operators.

\begin{figure}[ht]
    \centering
    \includegraphics[scale=2.8]{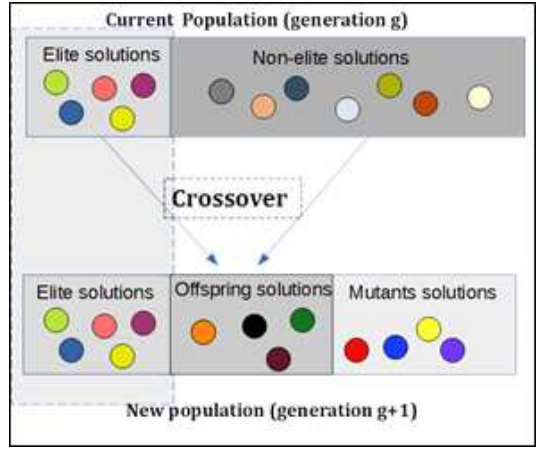}
    \caption{BRKGA application with the transition between generation $g$ and generation $g$+1}
    \label{fig:Fig1}
\end{figure}

The $p_m$ solutions produced from mutation are also copied to the population evaluated in the following generation. These solutions are generated analogously to the first generation of BRKGA. Finally, to complement the population evaluated in the next generation, $(p - p_e - p_m)$  solutions are produced by applying the crossover operator. Figure \ref{fig:Fig1} illustrates two generations in a row of BRKGA with the application of these operators, which is carried out during a fixed number of generations (maxGEN) of the BRKGA. Furthermore, the best solution is obtained at the end of all generations. For detailed information about this method, see \cite{Marti2018} and \cite{Goncalves2011}.

\subsection{STRATMH Heuristic} \label{STH} 

Aiming to find reasonable quality solutions for the stratification problem associated with optimization Problem (II), we present a heuristic algorithm based on the BRKGA method and the discretization presented in \ref{discre}. The proposed heuristic, called STRATMH, differs from the BRKGA in terms of the forms of representation and cons\-truction of the solutions that make up the population evaluated in each generation and the crossover operator.

\subsubsection{Representation and Construction of solutions} \label{PC}

 Each solution built by the heuristic is represented by a vector $w$ with $L$ values $w_h$, corresponding to the number of observations of $B$ in the $h^{th}$ stratum, such that the constraints (\ref{eq:n17}) and (\ref{eq:n18}), presented in \ref{discre} and here again, are fulfilled.
 \begin{equation*}
w_1 + \cdots + w_h + \cdots + w_L=|B|, 
\label{eq:n22}
\end{equation*}
\begin{equation*}
w_h \ge 2,  \ h =1,\dots,L.
\label{eq:n23}
\end{equation*}

 In order to ensure that every solution built satisfies these restrictions, the following procedure applies:
 \renewcommand{\theenumi}{\roman{enumi}}%
 \begin{enumerate}
     \item Select a random value $w_1 \in \{2, 3, \dots,|B|-2(L-1)\}$
      \item  Randomly select the number of observations from the $h^{th}$ stratum; that is, choose a random value $w_h$ $\in \{2, \dots ,(|B|-\sum_{i=1}^{h-1}w_i-2(L-h))\}$  for each $h = 2, \dots , L-1$
     \item Calculate the value of $w_L$ using  $|B|-\sum_{h=1}^{L-1} w_h$
\end{enumerate}     

 In this procedure, except for the last stratum, the choice number in the other strata is defined based on a random selection, considering the minimum (same) and maximum values per stratum. It is possible to verify that the use of (i), (ii), and (iii) guaranteed the fulfillment of (\ref{eq:n22})  and (\ref{eq:n23}). In the restriction in (\ref{eq:n22}), the fulfillment is given by the definition of step (iii). In the case of the restrictions in (\ref{eq:n23}), by the definition of (i) and by the recurrence relation in (ii), it is observed that each $w_h$ $(h=1,\dots,L-1)$  generated randomly will be greater than or equal to 2; missing to verify that $w_L \ge 2$   also occurs. Since $w_h \ge 2 \ (h=1,\dots,L-1)$, we have that $w_1+w_2+\dots+w_{L-1} \ge 2x(L-1)$. Finally, considering this last inequality and the hypothesis that   $|B| \ge 2L$  (as per subsection \ref{discre}), it has to $w_L=B-(w_1+...+w_L-1) \ge 2L-2(L-1) \ge 2$.
 
 Table \ref{tab:Table2} presents two feasible solutions concerning  (\ref{eq:n22})  and (\ref{eq:n23}) produced from applying the procedure mentioned previously, considering $L=4$ and $|B|=15$.
 
 \begin{table}[ht]
\tbl{Examples of Solutions  Produced}
{\begin{tabular}{ccc} \hline
Selected   $w_h$ values & Solution 1 $(w)$ & Solution 2 $(w)$  \\ \hline
$w_1 \in \{2,\dots,9\}$           & 5   & 3 \\
$w_2 \in \{2,\dots,15-w_1-4\}$     & 3   & 6 \\
$w_3 \in \{2,\dots,15-w_1-w_2-2\}$   & 4  & 3 \\
$w_4=|B|- w_1-w_2-w_3$ & 3 & 3 \\ \hline    
\end{tabular}}
\label{tab:Table2}
\end{table}

From this procedure, each $w$ vector generated corresponds to a possible $X_U$ stra\-tification. More specifically, considering each vector $w$, the values of $N_h$ and $S_{hx}^2$ are calculated, which are used in Equation (\ref{eq:n16}), together with the target cv ($cv_t$), in order to obtain the value of $n$, the objective function of the optimization model defined in Problem II. This procedure is used in the first generation (iteration) of the heuristic to produce the $p$ initial $w$ vectors (stratifications), and as a mutation operator, to produce, in the other generations, a quantity $p_m$ of new solutions, that is, vectors with possible stratifications.

\subsubsection{Crossover and Mutation Operators} \label{CMO}

In each generation, after calculating the objective function (sample size) for each of the $p$ vectors $w$, these vectors are arranged in a non-decreasing order. Then, an Elite set denoted by $S_e$  is defined, consisting of $p$ and $w$ vectors corresponding to the $p_e$ smallest sample sizes, with the ($p-p_e$) remaining vectors associated with the non-elite set denoted by $S_{ne}$. Finally, following the same principle as BRKGA, the elite set vectors, corresponding to the best stratifications, are added to the next generation's population, together with $p_m$ mutant vectors - corresponding to new stratifications obtained by applying the construction procedure described in \ref{PC}.

In order to obtain the $(p-p_e-p_m)$ remaining solutions evaluated in the next generation, is applied a crossover operator for each pair $(w_e, w_{ne})$, obtained from the random selection of a vector $w_e \in S_e$ and a vector $w_{ne} \in S_{ne}$. As for a crossed pair, two new vectors (stratifications) are produced; this operator must be applied $(p-p_e-p_m)/2$ times.

The crossover exchanges the values of each of the corresponding $L$ positions of $w_e$ and $w_{ne}$, redistributing the absolute difference $d$, where $d = |w_e[i]-w_{ne}[i]|  (i=1,\dots, L)$, between these values, proportionally to the values of $w_e$ and $w_{ne}$ associated with the other strata, adding or subtracting such difference by the other positions of this vector. At each exchange and redistribution, two new stratifications are produced for which the objective function's value is calculated. At the end of the $L$ exchanges, the two best solutions produced (with smaller sample sizes), among the $2L$ solutions, are added to the set formed by the $(p-p_e- p_m)$ solutions of the crossover.

\begin{figure}[ht]
    \centering
    \includegraphics[scale=1.1]{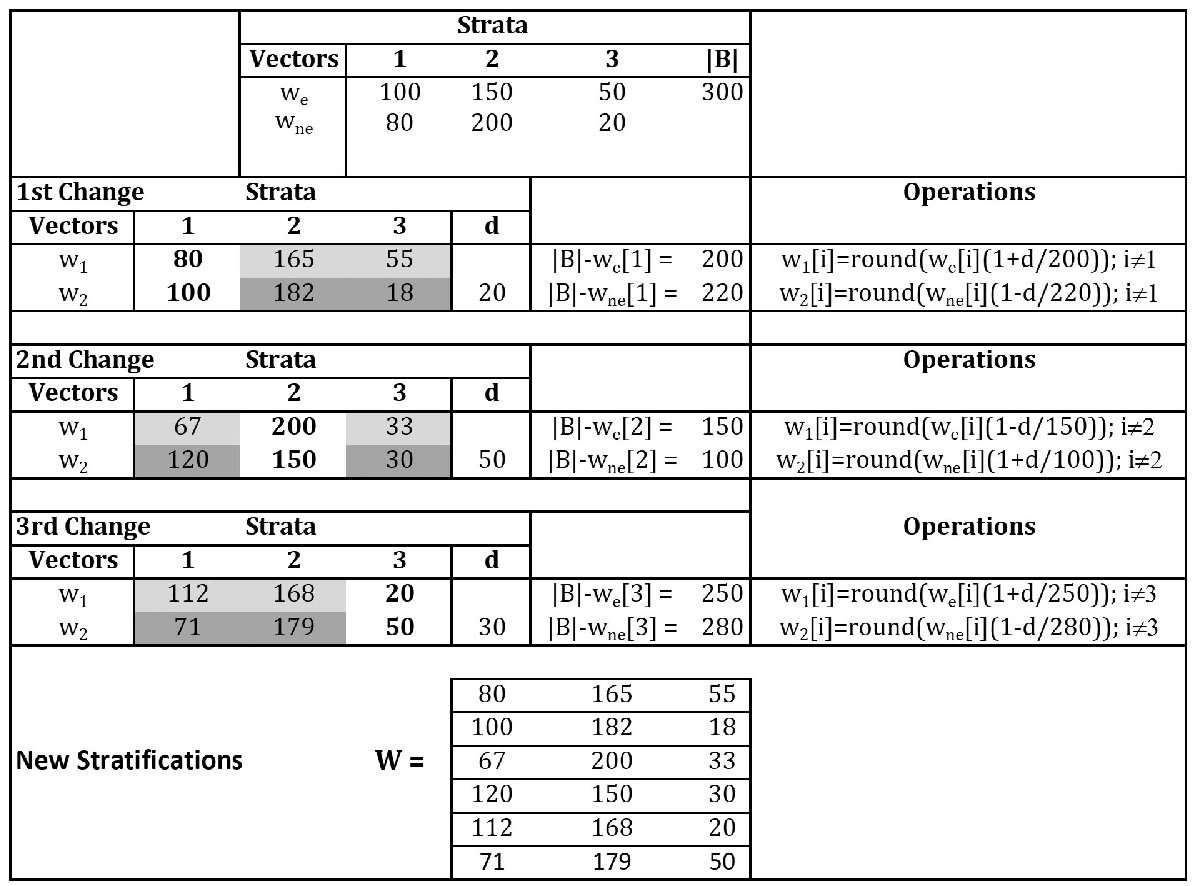}
    \caption{Crossover Operator Applied to an Example}
    \label{fig:Fig2}
\end{figure}

\begin{figure}[h]
 \centering
    \includegraphics[scale=1.3]{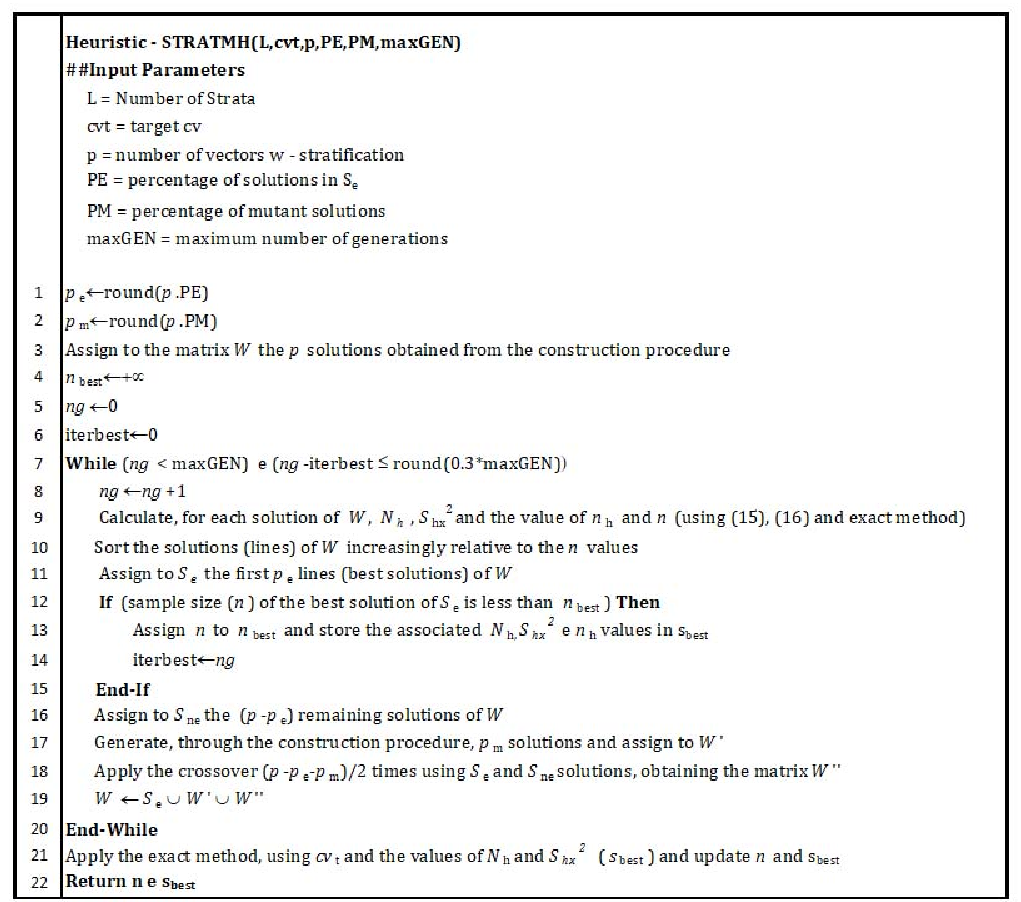}
    \caption{STRATMH heuristic pseudocode}
    \label{fig:Fig3}
\end{figure}

\begin{figure}[]
\centering
    \includegraphics[scale=1.9]{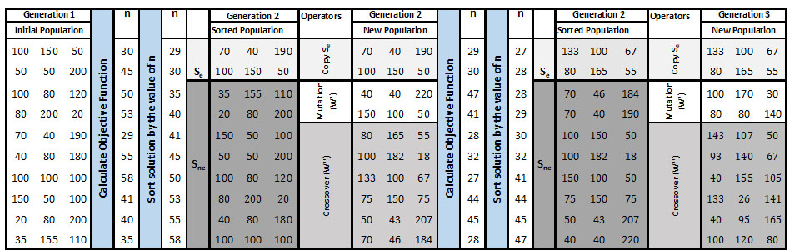}
    \caption{Illustration of the application of the STRATMH heuristic in three consecutive generations}
    \label{fig:Fig4}
\end{figure}

Figure \ref{fig:Fig2} illustrates the application of this operator in a pair of solutions $w_e$ and $w_{ne}$, considering $L=3$ and $|B|=300$. In Appendix 1, Figure \ref{fig:Fig7} presents the pseudocode that implements all the steps of this operator. Figure \ref{fig:Fig3} brings the complete pseudocode of the heuristic and Figure \ref{fig:Fig4} illustrates its application, considering the same values of $L$ and $|B|$ and $p=10$ (number of solutions - $w$ vectors).

About Figure \ref{fig:Fig3}, in lines 1 and 2, the percentages of elite solutions and solutions built on mutation are defined. In line 3, the initial $p$ solutions constructed are stored in a $W_{pxL}$ matrix. In each generation, between lines 7 and 20, new solutions are produced by evaluating the objective function (calculation of sample size) and applying the crossover and mutation operators. In line 9, the corresponding values of $N_h$, $S_{hx}^2$ are calculated for each line of $W$ ($w$ vector). Then, using equations (\ref{eq:n16}) and (\ref{eq:n15}), $n$ and $n_h$ are calculated and rounded. If the values of $n_h$ do not satisfy the constraints (\ref{eq:n12}) and (\ref{eq:n13})) of Problem (II), configuring an infeasibility, the exact method proposed in \cite{Brito2015} is applied to obtain integer $n$ and $n_h$; otherwise, rounded values are considered.

This method uses as an input data the values of $N_h$, $S_{hx}^2$ and $cv_t$, producing as a result, sample sizes $n_h (h=1,\dots, L)$ in which simultaneously minimize the objective function of Problem (II) and meet your restrictions. This method is applied in the case of infeasibility and in the best solution produced in the last generation of the heuristic because it significantly increases the computational cost of the heuristic if applied in all solutions.

To define  the sets $S_e$ and $S_{ne}$ (lines 11 and 16), the lines of $W$  are ordered about the values of $n$ obtained for each stratification. Between lines 12 and 15, the best update occurs solution to the problem. Then, to produce a new set of solutions evaluated in the population (matrix $W$) of the next generation, the crossover and mutation operators (lines 17 and 18) are used. Finally, considering the elite $S_e$ solutions (line 11) and the solutions 
$W$’ (mutation) and $W$’’ (crossover) obtained, respectively, the new population is defined (line 19).

The heuristic is performed until the maximum number of generations is reached (maxGEN) or until there is no reduction in the value of the objective function (sample size) for a pre-defined amount of generations (30\% of the total generations as identified in preliminary experiments). Finally, in line 21, the exact method proposed in \cite{Brito2015} is applied to the best solution obtained at the end of all generations.

\newpage

\section{Numerical Experiments} \label{NE}

This section presents the computational results where the STRATMH algorithm and four algorithms from the literature were evaluated: a modified version of the Lavallé and Hidiroglou (LH) algorithm \cite{Rivest2002}, a version of the algorithm (Random Search) proposed by Kozak (KO) \cite{Kozak2004}, VNS algorithm \cite{Brito2021} and genetic algorithm proposed by Balin and Barcarolli (BB) \cite{Ballin2013}. These four algorithms are implemented and available in R packages, and the STRATMH algorithm, implemented in R, is available in GitHub (stratMH function), as shown in Table \ref{tab:Table3}. In addition, Appendix 2 brings examples of the use of STRATMH.

 \begin{table}[ht]
\tbl{Information about functions used in Algorithms}
{\begin{tabular}{c|ccc} \hline
Algorithm/Reference & Name & Package or Repository & Function Name   \\ \hline
Lavallé and Hidiroglou \cite{Rivest2002}   & LH        & stratification   & strata.LH  \\
Kozak {[}11{]} & KO & stratification        & strata.LH           \\
Brito et al \cite{Brito2021}    & VNS  & stratvns    & STRATVNS   \\
Ballin and Barcarolli \cite{Ballin2013}   & BB   & SamplingStrata & optimStrata    \\
Proposed Heurisctic & STRATMH     & https://github.com/jambrito/STRATMH    & stratMH    \\  \hline
\end{tabular}}
\label{tab:Table3}
\end{table}

The remainder of this section provides information about the populations used in the experiments, and subsection \ref{CR} presents the results and analysis of the experiments.

In the experiments reported in this section, 31 populations described in Table \ref{tab:Table3} were used, 28 of them from the literature used in \cite{Kozak2006}, \cite{Brito2021, Brito2017, Brito2019} and \cite{Ballin2013}, and available in the packages of the R language: GA4Stratification (although removed from the CRAN, can be found at https://cran.r-project.org/src/contrib/Archive/GA4Stratification/), stratification \cite{BR2011}, stratifyR \cite{RK2020}, and sampling (available on the R CRAN). In addition, the authors of this article generated three new populations using functions from the EnvStats package.

Table \ref{tab:Table14} presents information on populations and Table \ref{tab:Table4} presents some associated descriptive statistics, such as name, population size ($N$), the cardinality of $B$, minimum and maximum values for variable $X$, and the value of its asymmetry coefficient ($AS$).

\begin{table}[]
\tbl{Description of the 31 populations }
{
\resizebox{\textwidth}{!}{%
\begin{tabular}{lcl} \hline
Names                     & Reference                        & Description                                                                                      \\ \hline
AgrMinas*  & \cite{Brito2019} & Agricultural production of municipalities in Minas Gerais State, Brazil, from 2006.\\
Anaemia\_Haemoglobin & stratifyR & The Anaemia data comes from the Fiji National Nutritional Survey in   2004        \\
BeeFarms    & \cite{Brito2017,Brito2019,Brito2021,Ballin2013}  & Australian cattle farms stratified by industrial regions   \\
beta103   & GA4Stratification   & Generated from Beta distribution with parameters $a= 10$ and $b=3$ \\
CensoCO*      & \cite{Brito2019} & Data from the   2012 school census in Brazil for the mid-west region\\
Chi1          & GA4Stratification        & Generated by the Chi-Square distribution with 1 degree of freedom    \\
Chi5     &  GA4Stratification        & Generated by the Chi-Square distribution with 5 degree of freedom\\
Debtors*                  & stratification           & Debtor population in an Irish firm   \\
GAMMA2000                 & EnvStats    & Generated by the gamma distribution - rgamma(2000,2,0.5)     \\
hies\_income & stratifyR   & Household Income Expenditure Survey (HIES) in Fiji  in the year 2010\\
iso2004                   & GA4Stratification        & Net sales of Turkish industrial companies in 2004.    \\
Kozak1, Kozak2, Kozak3*   & \cite{Kozak2006} \cite{Brito2017} \cite{Brito2019} \cite{Brito2021} & Generated based   on X=exp(z), where Z is a realization of a normal random   variable.       \\
ME84    & sampling     & Number of municipal employees in 284 municipalities in Sweden in 1984    \\
MRTS*  & stratification & Population   simulated from the Monthly Survey on Sales in Retail Trade from Statistics Canada\\
p100e10 & GA4Stratification & Generated by the normal distribution with $\mu$=100 and $\sigma$=10  \\
p75 & GA4Stratification & Population in thousands of 284 municipalities in Sweden in 1975  \\
pop1076, pop1616, pop2911 & \cite{Brito2017} \cite{Brito2019} & Extracted from the Brazilian Annual Manufacturing   Survey – Number of employees \\
pop500,  pop800 &  \cite{Brito2019}, \cite{Brito2021}    & Generated by the log-normal distribution  $X=e^Z$   \\
& & where $Z$ is normal with $\mu$=4 and $\sigma^2=2.7$  \\
REV84  & sampling     & Property values in millions of Swedish kronor from 284 municipalities   in 1984                  \\
SugarCaneFarms* & \cite{Brito2017},\cite{Brito2019}, \cite{Brito2021},\cite{Ballin2013} & Australia's sugar cane farm population \\
Swiss                     & sampling     & Information on Swiss municipalities (2003)  \\
TRI900  & EnvStats & Generated by the gamma distribution - rtri(900, min = 0, max = 1, mode   = 1/2)  \\
Usbanks     & stratification           & Million-dollar funds from major US commercial banks    \\
Uscities & stratification  & Population in thousands of American cities in 1940 \\
Uscolleges  & stratification  & Number of students at four-year US colleges in 1952-1953 \\
WEIBULL1500*    & EnvStats & Generated by the weibull distribution - rweibull(1500,0.5)  \\ \hline         
\end{tabular}
}}
\tabnote{\textsuperscript{*} Populations used in the experiments to calibrate the algorithm in subsection \ref{CR}}
\label{tab:Table14}
\end{table}

\begin{table}[htbp]
\tbl{Basic information of the 31 populations}
{
\tiny
\begin{tabular}{l|rrrrr} \hline
Name                 & $N$     & $|B|$   & Min        & Max       & AS       \\ \hline
AgrMinas*             & 844   & 226   & 5          & 47,800    & 7.30877  \\
anaemia\_Haemoglobin & 724   & 90    & 6.1        & 16.9      & -0.43629 \\
BeefFarms            & 430   & 353   & 50         & 24250     & 4.56352  \\
beta103              & 1,000 & 1,000 & 357.983    & 985.9574  & -0.69695 \\
CensoCO*             & 9,977 & 79    & 1          & 911       & 40.03439 \\
chi1                 & 1,000 & 1,000 & 4.23E-06   & 12.98871  & 2.66591  \\
chi5.txt             & 1,000 & 1,000 & 0.059874   & 23.42703  & 1.40219  \\
Debtors*              & 3,369 & 1,129 & 40         & 28,000    & 6.43743  \\
GAMMA2000            & 2,000 & 2,000 & 0.056305   & 16.22303  & 1.27668  \\
hies\_income         & 3,566 & 3,491 & 34.1224    & 132921.6  & 2.88252  \\
iso2004              & 487   & 487   & 63,582,908 & 1.04E+10  & 10.02768 \\
Kozak1               & 4,000 & 51    & 3          & 72        & 1.39607  \\
Kozak2               & 4,000 & 2837  & 243        & 28,578    & 2.65563  \\
Kozak3*               & 2,000 & 581   & 6          & 2,793     & 3.54417  \\
ME84                 & 284   & 264   & 173        & 47,074    & 8.63944  \\
MRTS*                & 2,000 & 2,000 & 141.2      & 486,366.5 & 8.60937  \\
p100e10              & 1,000 & 1000  & 73.55778   & 127.3164  & -0.03064 \\
p75                  & 284   & 68    & 4          & 671       & 8.42744  \\
pop1076              & 1,076 & 88    & 5          & 1,643     & 13.23037 \\
pop1616              & 1,616 & 165   & 5          & 2,618     & 11.09524 \\
pop2911              & 2,911 & 247   & 5          & 2,497     & 11.49960 \\
pop500               & 500   & 261   & 1          & 4,784,142 & 21.53249 \\
pop800               & 800   & 402   & 1          & 473510    & 22.12974 \\
rev84                & 284   & 277   & 347        & 59,877    & 7.83394  \\
SugarCaneFarms*      & 338   & 101   & 18         & 280       & 2.26478  \\
Swiss                & 2,896 & 881   & 0          & 3,634     & 2.72729  \\
TRI900               & 900   & 900   & 0.003221   & 0.97282   & -0.01988 \\
USbanks              & 357   & 200   & 70         & 977       & 2.06696  \\
UScities             & 1,038 & 116   & 10         & 198       & 2.86826  \\
UScolleges           & 677   & 576   & 200        & 9,623     & 2.45168  \\
WEIBULL1500*          & 1,500 & 1,500 & 3.68E-08   & 38.25109  & 4.32759 \\ \hline
\end{tabular}
}
\tabnote{\textsuperscript{*} Populations used in the experiments to calibrate the algorithm in subsection \ref{CR}}
\label{tab:Table4}
\end{table}

\newpage

\subsection{Computational Results} \label{CR}

The experiments with the five algorithms were carried out in RStudio, using a computer with 16GB of RAM and equipped with an AMD FX-6300 3.5 GHz processor. Regarding the functions presented in Table \ref{tab:Table3}, associated with the four algorithms in the literature, they have used the default values of its parameters.

In the case of the parameters used in the STRATMH heuristic, a preliminary calibration experiment was carried out, based on the analyzes and recommendations made in \cite{Goncalves2011} and taking as a reference a set of values for each of its parameters, as shown in Table  \ref{tab:Table5}. Such an experiment is essential since the value considered in the set of parameters is a factor that contributes significantly to the good performance of any algorithm based on a stochastic optimization method.

To determine the ideal combination of the parameters of the proposed heuristic, seven populations were used in this experiment, which we marked with an asterisk in Table \ref{tab:Table4}. The heuristic was performed 10 times to stratify each of these populations, for each $L \in \{3,4,5,6,7\}$, $cv_t=5\%$ and $cv_t=10\%$ and considering 320 combinations of the parameters $p$, maxGEN, $p_e$ , $p_m$ (Table \ref{tab:Table5}) – making, per population, 32,000 executions (number of strata x target cv x combinations of parameters x number of executions), being stored, by execution, the final value of the objective function, that is, smaller sample size.

\begin{table}[htbp]
\tbl{STRATMH parameters for calibration experiments}
{
\small
\begin{tabular}{l|ll} \hline
Parameter & Description                     & Values                      \\ \hline
$p$  & size of population  & 20, 30, 40, 50,   100     \\
& (number of $w$ vectors used by generation) & \\
maxGEN    & Number of   generations  & 20, 30, 40, 50              \\
$p_e$        & size of elite   population       & 0.15$p$,   0.2$p$, 0.25$p$, 0.30$p$ \\
$p_m$        & size of mutant   population   & 0.15$p$,   0.2$p$, 0.25$p$, 0.30$p$ \\ \hline
\end{tabular}}
\label{tab:Table5}
\end{table}

\newpage

In the first stage of the experiment, considering each population, the number of strata, and target cv, the best solution was the one associated with the smallest sample size (best solution) obtained in the 3,200 runs (combination of parameters x 10 trials). Then, the ratio between the number of runs that each combination produced a sample size equal to the best solution and the total runs of the heuristic using that combination was calculated - in case 10. The closer to 1 such ratio, the more times the combination produced the best solution.

Then, considering such proportions for each of the populations x number of strata x target cv's (7 x 5 x 2), the proportions were decreasingly ordered, taking, by po\-pulation, number of strata ($L$) and target cv's ($cv_t$), the combinations of parameters associated with the first five proportions in such ordering – making a list of 350 combinations. Finally, considering all the combinations, the most frequent combination in this list was chosen as the winning combination, reaching the combination: $p$=50, maxGEN=50, $p_e=0.3p=15$, and $p_m=0.3p=15$.

Considering this combination for the STRATMH algorithm and the default para\-meters of the four algorithms in the literature, the five algorithms were applied in each of the 31 populations (Table \ref{tab:Table4}), being evaluated 20 different scenarios corresponding to the number of strata ($L$) between 3 and 7 and target cv's ($cv_t$) of, respectively, 3\%, 5\%, 7.5\%, and 10\%, making a total of 620 solutions. In addition, their sample sizes and processing times were evaluated for comparison and analysis of the algorithms.

It was observed that the sample size produced, about 31\% of the solutions, when evaluating the solutions produced by the BB algorithm, concerning some populations and scenarios, corresponded to a lower number of strata than that previously esta\-blished; a fact resulting from this algorithm to favor, in the stratification, the smallest possible sample size to the detriment of the number of strata. Additionally, in the case of the LH algorithm, considering the solutions produced for some populations and scenarios, non-convergence was observed in about 5\% of the solutions. More specifically, the algorithm produced a sample size whose associated coefficient of variation was greater than the target cv ($cv_t$).

From these observations, in order to enable a correct comparison between the five algorithms, three types of analysis were performed: (i) the first analysis is performed using the 620 solutions produced by the KO, VNS, and STRAMH algorithms; (ii) the second, consider the intersection of the set of 587 valid solutions produced by the LH algorithm and the 620 solutions produced by the KO, VNS and STRATMH algorithms; (iii) the third considers the analysis of the five algorithms, considering the intersection of sets of valid solutions (total of 396) produced by the BB and LH algorithm together with the other algorithms, that is, the cases in which the number of strata produced by the BB algorithm corresponds to the number previously fixed and that the LH algorithm converged.

\subsubsection{Analysis I - KO, VNS and STRATMH Algorithms}

Table \ref{tab:Table6} brings the percentage of best solutions for the three algorithms by the number of strata and target cv’s, and Table \ref{tab:Table7} brings the overall percentage of best solutions for each algorithm. Here, the best solution is the one corresponding to the smallest sample size. From these tables, it is observed that the STRATMH algorithm performed better than the KO and VNS algorithms, with percentages of better solutions greater than or equal (in bold) to these two algorithms in 19 of the 20 analyzed scenarios, with a percentage lower only than that of the algorithm VNS in the scenario associated with $L=7$ and $cv_t$=10\%. Considering the overall percentage of solutions (Table \ref{tab:Table7}), STRAMH reached 96.0\%, followed by the VNS algorithms with 91.0\% and KO with 74.4\%.

\begin{table}[htbp]
\tbl{Percentage of  best solutions produced by algorithm, number of strata and $cv_t$}
{
\resizebox{\textwidth}{!}{%
\begin{tabular}{l|rrrrrrrrrr} \hline
 & \multicolumn{1}{l}{}  & \multicolumn{2}{l}{$cv_t$=3.0\%}  & \multicolumn{1}{l}{}  & \multicolumn{1}{l}{}  & \multicolumn{1}{l}{}  & \multicolumn{2}{l}{$cv_t$=5.0\%}                          & \multicolumn{1}{l}{}  & \multicolumn{1}{l}{}  \\ \hline
 & \multicolumn{1}{c}{}  & \multicolumn{1}{c}{}          & \multicolumn{1}{c}{$L$} & \multicolumn{1}{c}{}  & \multicolumn{1}{c}{}  & \multicolumn{1}{c}{}  & \multicolumn{1}{c}{}           & \multicolumn{1}{c}{$L$} & \multicolumn{1}{c}{}  & \multicolumn{1}{c}{}  \\
Algorithm & \multicolumn{1}{c}{3} & \multicolumn{1}{c}{4}         & \multicolumn{1}{c}{5} & \multicolumn{1}{c}{6} & \multicolumn{1}{c}{7} & \multicolumn{1}{c}{3} & \multicolumn{1}{c}{4}          & \multicolumn{1}{c}{5} & \multicolumn{1}{c}{6} & \multicolumn{1}{c}{7} \\ \hline
KO        & 87.1     & 71.0  & 67.7   & 64.5   & 35.5   & 83.9 & 87.1 & 67.7 & 74.2  & 51.6 \\
STRATMH   & \textbf{100.0} & \textbf{93.5}  & \textbf{93.5} & \textbf{96.8}  & \textbf{80.6} & \textbf{96.8} & \textbf{100.0}   & \textbf{96.8}         & \textbf{90.3}         & \textbf{96.8}\\
VNS       & \textbf{100.0} & 90.3  & 90.3  & 80.6 & 64.5 & \textbf{96.8}  & 87.1 & 90.3  & \textbf{90.3} & 77.4  \\ \hline
          & \multicolumn{1}{c}{}  & \multicolumn{1}{c}{$cv_t$=7.5\%} & \multicolumn{1}{c}{}  & \multicolumn{1}{c}{}  & \multicolumn{1}{c}{}  & \multicolumn{1}{c}{}  & \multicolumn{1}{c}{$cv_t$=10.0\%} & \multicolumn{1}{c}{}  & \multicolumn{1}{c}{}  & \multicolumn{1}{c}{}  \\ \hline
          & \multicolumn{1}{c}{}  & \multicolumn{1}{c}{}          & \multicolumn{1}{c}{$L$} & \multicolumn{1}{c}{}  & \multicolumn{1}{c}{}  & \multicolumn{1}{c}{}  & \multicolumn{1}{c}{}           & \multicolumn{1}{c}{$L$} & \multicolumn{1}{c}{}  & \multicolumn{1}{c}{}  \\
Algorithm & \multicolumn{1}{c}{3} & \multicolumn{1}{c}{4}         & \multicolumn{1}{c}{5} & \multicolumn{1}{c}{6} & \multicolumn{1}{c}{7} & \multicolumn{1}{c}{3} & \multicolumn{1}{c}{4}          & \multicolumn{1}{c}{5} & \multicolumn{1}{c}{6} & \multicolumn{1}{c}{7} \\ \hline
KO    & 87.1 & 64.5  & 64.5  & 74.2  & 83.9  & 77.4  & 74.2  & 87.1  & 87.1  & 96.8                  \\
STRATMH   & \textbf{100.0} & \textbf{100.0}  & \textbf{93.5} & \textbf{96.8}    & \textbf{96.8} & \textbf{100.0} & \textbf{100.0}  & \textbf{96.8} & \textbf{96.8} & 93.5                  \\
VNS       & \textbf{100.0}  & 96.8 & 87.1  & 93.5  & 87.1  & \textbf{100.0} & 96.8 & 93.5 & \textbf{96.8}  & \textbf{100.0} \\      \hline
\end{tabular}}
}
\label{tab:Table6}
\end{table}

\begin{table}[htbp]
\tbl{Percentage of times that algorithm produced the best solution}
{
\small
\begin{tabular}{l|r} \hline
Algorithm & \% Times best \\  \hline
STRATMH     & 96.0  \\
VNS         & 91.0  \\
KO          & 74.4  \\  \hline
\end{tabular}}
\label{tab:Table7}
\end{table}

In addition to the percentages of best solutions, the relative efficiency ($Eff$) of the STRATMH algorithm was evaluated about the VNS and KO algorithms, that is, the ratio between the sample sizes produced. Considering each of the 620 sample sizes (solutions) obtained, the value of expressions in (\ref{eq:n24}) and (\ref{eq:n25}) was calculated for each population and scenario ($L$ and $cv_t$), with $j$ being the index corresponding to the $j^{th}$ solution and $n^{KO}_j$ , $n^{VNS}_j$ e $n^{STRATMH}_j$ correspond to the minimum sample sizes to reach the predefined target cv, considering $L$ strata and the KO, VNS, and STRATMH algorithms. The higher the $Eff$ value, the more significant the difference between the sample sizes produced by the STRATMH algorithm relative to these algorithms.

\begin{equation}
Eff(KO,STRATMH)=\frac{n_j^{KO}}{n_j^{STRATMH}}, \; \; j=1,\dots,620
\label{eq:n24}
\end{equation}

\begin{equation}
Eff(VNS,STRATMH)=\frac{n_j^{VNS}}{n_j^{STRATMH}}  ,  \; \; j=1,\dots,620.
\label{eq:n25}
\end{equation}

\begin{figure}[]
    \centering
    \includegraphics[scale=0.5]{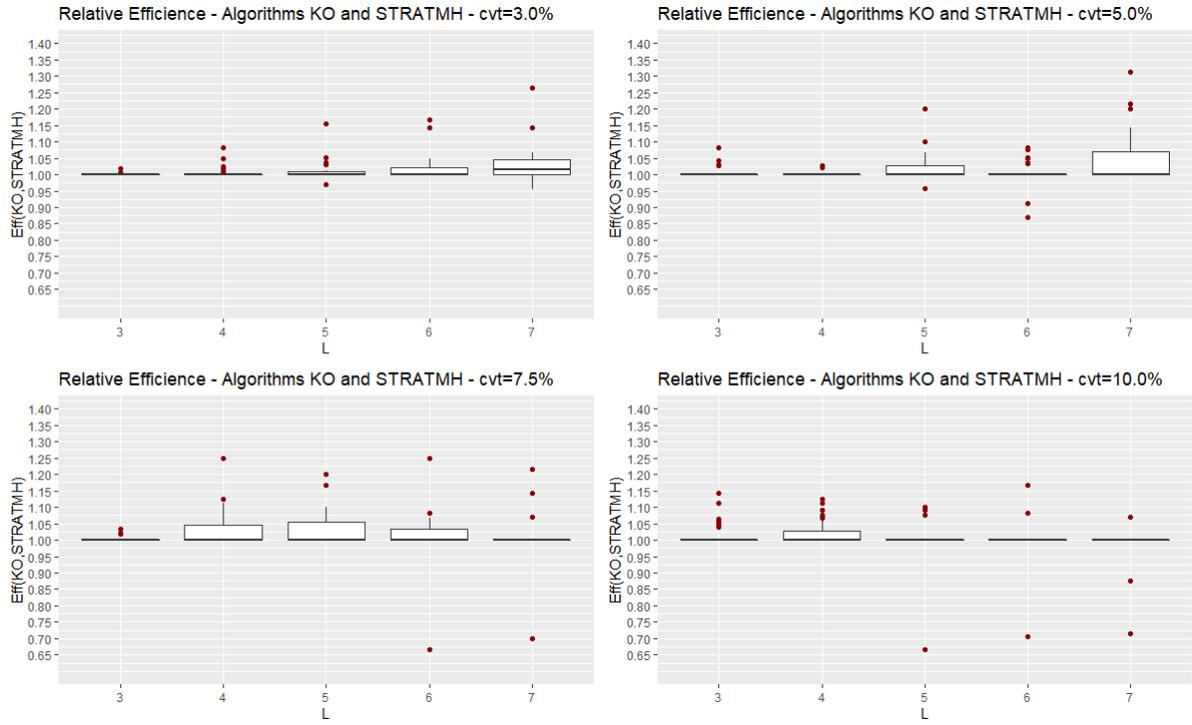}
    \caption{Relative Efficiences by strata and target cv – Algorithms KO and STRATMH}
    \label{fig:Fig5}
\end{figure}

\begin{figure}[]
\centering
\includegraphics[scale=0.5]{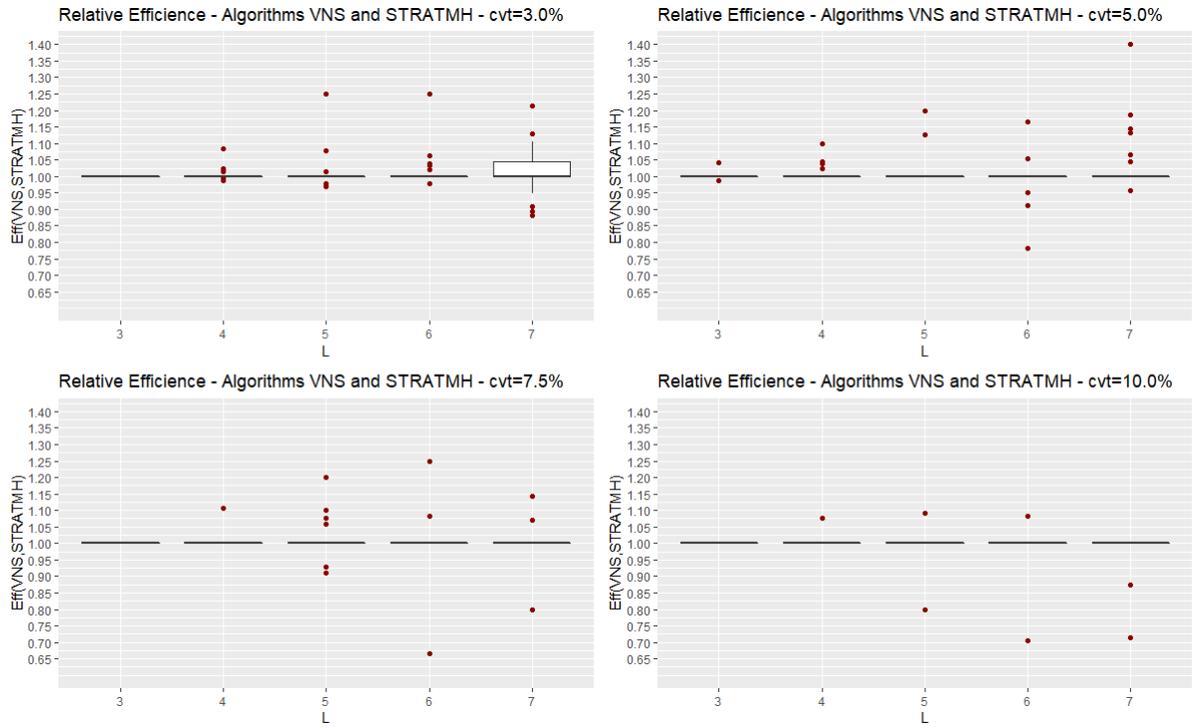}
\caption{Relative Efficiences by strata and target cv – Algorithms VNS and STRATMH}
\label{fig:Fig6}
\end{figure}

These values were used to build the boxplots presented in Figures \ref{fig:Fig5} and \ref{fig:Fig6}, which show the distributions of the relative efficiencies of the STRATMH algorithm about the KO and VNS algorithms by the number of strata and coefficients of target variation. From the analysis of these figures, it is observed that the STRATMH algorithm's most significant gains were concerning the KO algorithm. Furthermore, the STRATMH algorithm, in most cases, has a relative efficiency better than or equal to these algorithms. Finally, the smaller the value of $cv_t$, that is, when a higher level of precision is required, the greater the gains of the STRATMH algorithm.

\subsubsection{Analysis II - KO, LH, VNS and STRATMH Algorithms}

Table \ref{tab:Table8}  brings, by the number of strata and target cv's, the percentage of best solutions concerning the sample sizes produced by the KO, LH, VNS, and STRATMH algorithms, and Table \ref{tab:Table9} brings the overall percentage of each of the algorithms, considering all the 587 valid solutions produced by the LH algorithm. From \ref{tab:Table8}, it can be seen that, in 18 of the 20 scenarios (in bold), STRATMH produced percentages greater than or equal to those of the other algorithms, followed by the VNS and KO algorithms. The only scenarios where STRATMH performed less than VNS correspond to $L$=6, 7, and cvt=10\%. Additionally, the algorithm produced, in 18 of the 20 scenarios, a percentage of the best solutions above 93\%, with only a percentage of the order of 79\% ($L=7$ and $cv_t$=3\%) and a percentage of the order of 90 being observed \% ($L$=6 and $cv_t$=5\%). Furthermore, compared to the second-best algorithm (VNS), STRATMH produced the highest percentages of best solutions in 12 of the 20 scenarios. Finally, the biggest gains of this algorithm about the other algorithms, translated in percentage terms, are observed for $cv_t$=7.5\%.

\begin{table}[htbp]
\tbl{Percentage of  best solutions produced by algorithm. number of strata and $cv_t$}
{
\resizebox{\textwidth}{!}{%
\begin{tabular}{l|rrrrrrrrrr} \hline
 & \multicolumn{1}{l}{}  & \multicolumn{2}{l}{$cv_t$=3.0\%}  & \multicolumn{1}{l}{}  & \multicolumn{1}{l}{}  & \multicolumn{1}{l}{}  & \multicolumn{2}{l}{$cv_t$=5.0\%}                          & \multicolumn{1}{l}{}  & \multicolumn{1}{l}{}  \\ \hline
 & \multicolumn{1}{c}{}  & \multicolumn{1}{c}{}          & \multicolumn{1}{c}{$L$} & \multicolumn{1}{c}{}  & \multicolumn{1}{c}{}  & \multicolumn{1}{c}{}  & \multicolumn{1}{c}{}           & \multicolumn{1}{c}{$L$} & \multicolumn{1}{c}{}  & \multicolumn{1}{c}{}  \\
Algorithm & \multicolumn{1}{c}{3} & \multicolumn{1}{c}{4}         & \multicolumn{1}{c}{5} & \multicolumn{1}{c}{6} & \multicolumn{1}{c}{7} & \multicolumn{1}{c}{3} & \multicolumn{1}{c}{4}          & \multicolumn{1}{c}{5} & \multicolumn{1}{c}{6} & \multicolumn{1}{c}{7} \\ \hline

LH     & 33.3   & 21.4 & 10.3& 15.4  & 14.3    & 33.3    & 21.4  & 14.3  & 24.1 & 37.9 \\
KO     & 86.7   & 67.9  & 65.5& 69.2  & 32.1    & 83.3    & 85.7  & 64.3  & 75.9 & 51.7 \\
STRATMH& \textbf{100.0}  & \textbf{92.9}  & \textbf{93.1} & \textbf{96.2} & \textbf{78.6} & \textbf{96.7}    & \textbf{100.0} & \textbf{96.4}  & \textbf{89.7} & \textbf{96.6} \\
VNS    & \textbf{100.0}  & 89.3 & \textbf{89.7} & 84.6  & 71.4 & \textbf{96.7} & 85.7  & 89.3  & 89.7 & 79.3 \\ \hline
Npops*	& 30	& 28	& 29	& 26	& 28	& 30	& 28	& 28	& 29	& 29 \\\hline

 & \multicolumn{1}{c}{}  & \multicolumn{1}{c}{$cv_t$=7.5\%} & \multicolumn{1}{c}{}  & \multicolumn{1}{c}{}  & \multicolumn{1}{c}{}  & \multicolumn{1}{c}{}  & \multicolumn{1}{c}{$cv_t$=10.0\%} & \multicolumn{1}{c}{}  & \multicolumn{1}{c}{}  & \multicolumn{1}{c}{}  \\ \hline
          & \multicolumn{1}{c}{}  & \multicolumn{1}{c}{}          & \multicolumn{1}{c}{$L$} & \multicolumn{1}{c}{}  & \multicolumn{1}{c}{}  & \multicolumn{1}{c}{}  & \multicolumn{1}{c}{}           & \multicolumn{1}{c}{$L$} & \multicolumn{1}{c}{}  & \multicolumn{1}{c}{}  \\
Algorithm & \multicolumn{1}{c}{3} & \multicolumn{1}{c}{4}         & \multicolumn{1}{c}{5} & \multicolumn{1}{c}{6} & \multicolumn{1}{c}{7} & \multicolumn{1}{c}{3} & \multicolumn{1}{c}{4}          & \multicolumn{1}{c}{5} & \multicolumn{1}{c}{6} & \multicolumn{1}{c}{7} \\ \hline
LH     & 29.0   & 23.3 & 40.0 & 43.3 & 53.3 & 32.3 & 36.7  & 63.3 & 73.3 & 76.7 \\
KO     & 87.1   & 63.3 & 66.7 & 76.7 & 86.7 & 77.4 & 76.7  & 90.0 & 90.0 & 96.7 \\
STRATMH& \textbf{100.0}  & \textbf{100.0} & \textbf{93.3} & \textbf{96.7} & \textbf{96.7} & \textbf{100.0} & \textbf{100.0} & \textbf{96.7} & 96.7 & 96.7 \\
VNS    & 100.0  & 96.7 & 86.7 & 96.7 & 86.7 & 100.0& 96.7  & 93.3 & 100.0& 100.0\\ \hline 
Npops*	& 31	& 30	& 30	& 30	& 30	& 31	& 30	& 30	& 30	& 30\\ \hline
\end{tabular}}
}
\tabnote{\textsuperscript{*} Number of populations where LH algorithm converged }
\label{tab:Table8}
\end{table}

In Table \ref{tab:Table9}, note that STRATMH has superior gain with an overall percentage of the order of 96.0\% of the best solutions, followed by the VNS algorithms with around 92.0\% and KO with 75.0\%.

\begin{table}[ht]
\tbl{Percentage of times that algorithm produced the best solution}
{
\small
\begin{tabular}{l|r} \hline
Algorithm & \% Times best \\  \hline
STRATMH	& 95.9 \\
VNS	& 91.8 \\
KO	& 75.0\\ 
LH	& 35.3\\ \hline
\end{tabular}
}
\label{tab:Table9}
\end{table}

\newpage

\subsubsection{Analysis III – KO, LH, VNS, BB and STRATMH Algorithms}

In this last analysis, five algorithms were evaluated considering only the feasible solutions produced by the BB and LH algorithms (a total of 396). From the analysis of Table \ref{tab:Table10}, which brings the percentage of best solutions per algorithm, the number of strata, and $cv_t$ , and \ref{tab:Table11}, which brings the overall percentage of best solutions for each of the algorithms, it is observed that the STRATMH algorithm presented, again, superior performance to other algorithms. Table \ref{tab:Table10} shows, in 18 of the scenarios (in bold), percentages of best solutions (greater than or equal to the other algorithms), with STRATMH showing a lower percentage only in the scenario associated with $L$=6 and $cv_t$=5\%, when compared to VNS, which had the second-best performance. Regarding the overall percentage of best solutions Table (\ref{tab:Table11}), the STRATMH algorithm had again a percentage above 94\%, followed by the VNS (89.1\%), KO (64.1\%), BB (51.5\%), and LH(15.7\%) algorithms.

\begin{table}[htbp]
\tbl{Percentage of  best solutions produced by algorithm, number of strata and $cv_t$}
{
\resizebox{\textwidth}{!}{%
\begin{tabular}{l|rrrrrrrrrr} \hline
 & \multicolumn{1}{l}{}  & \multicolumn{2}{l}{$cv_t$=3.0\%}  & \multicolumn{1}{l}{}  & \multicolumn{1}{l}{}  & \multicolumn{1}{l}{}  & \multicolumn{2}{l}{$cv_t$=5.0\%}                          & \multicolumn{1}{l}{}  & \multicolumn{1}{l}{}  \\ \hline
 & \multicolumn{1}{c}{}  & \multicolumn{1}{c}{}          & \multicolumn{1}{c}{$L$} & \multicolumn{1}{c}{}  & \multicolumn{1}{c}{}  & \multicolumn{1}{c}{}  & \multicolumn{1}{c}{}           & \multicolumn{1}{c}{$L$} & \multicolumn{1}{c}{}  & \multicolumn{1}{c}{}  \\
Algorithm & \multicolumn{1}{c}{3} & \multicolumn{1}{c}{4}         & \multicolumn{1}{c}{5} & \multicolumn{1}{c}{6} & \multicolumn{1}{c}{7} & \multicolumn{1}{c}{3} & \multicolumn{1}{c}{4}          & \multicolumn{1}{c}{5} & \multicolumn{1}{c}{6} & \multicolumn{1}{c}{7} \\ \hline
BB	& 44.8	& 32.0	& 15.4	& 13.6	& 17.4	& 63.0	& 60.0	& 40.9	& 22.2	& 35.3 \\
KO	& 82.8	& 64.0	& 61.5	& 68.2	& 21.7	& 77.8	& 84.0	& 59.1	& 61.1	& 29.4 \\
LH	& 27.6	& 12.0	& 0.0	& 4.5	& 0.0	& 25.9	& 12.0	& 0.0	& 11.1	& 17.6 \\
STRATMH	& \textbf{96.6}	& \textbf{92.0}	& \textbf{92.3}	& \textbf{95.5}	& \textbf{73.9}	& \textbf{92.6}	& \textbf{100.0}	& \textbf{95.5}	& 83.3	& \textbf{94.1} \\
VNS	& \textbf{96.6}	& 88.0	& 88.5	& 86.4	& 69.6	& \textbf{92.6}	& 84.0	& 90.9	& \textbf{88.9}	& 70.6 \\ \hline
Npops*	& 29	& 25	& 26	& 22	& 23	& 27	& 25	& 22	& 18	& 17 \\ \hline

 & \multicolumn{1}{c}{}  & \multicolumn{1}{c}{$cv_t$=7.5\%} & \multicolumn{1}{c}{}  & \multicolumn{1}{c}{}  & \multicolumn{1}{c}{}  & \multicolumn{1}{c}{}  & \multicolumn{1}{c}{$cv_t$=10.0\%} & \multicolumn{1}{c}{}  & \multicolumn{1}{c}{}  & \multicolumn{1}{c}{}  \\ \hline
          & \multicolumn{1}{c}{}  & \multicolumn{1}{c}{}          & \multicolumn{1}{c}{$L$} & \multicolumn{1}{c}{}  & \multicolumn{1}{c}{}  & \multicolumn{1}{c}{}  & \multicolumn{1}{c}{}           & \multicolumn{1}{c}{$L$} & \multicolumn{1}{c}{}  & \multicolumn{1}{c}{}  \\
Algorithm & \multicolumn{1}{c}{3} & \multicolumn{1}{c}{4}         & \multicolumn{1}{c}{5} & \multicolumn{1}{c}{6} & \multicolumn{1}{c}{7} & \multicolumn{1}{c}{3} & \multicolumn{1}{c}{4}          & \multicolumn{1}{c}{5} & \multicolumn{1}{c}{6} & \multicolumn{1}{c}{7} \\ \hline
BB	& 78.6	& 80.8	& 42.9	& 46.2	& 80.0	& 92.6	& 81.8	& 85.7	& 66.7 &	- \\
KO	& 82.1	& 57.7	& 52.4	& 46.2	& 40.0	& 74.1	& 63.6	& 78.6	& 83.3 &	- \\
LH	& 17.9	& 11.5	& 23.8	& 7.7	& 0.0	& 22.2	& 18.2	& 50.0	& 66.7 &	- \\
STRATMH	& \textbf{96.4} &	\textbf{100.0}	& \textbf{90.5}	& \textbf{100.0}	& \textbf{100.0}	& \textbf{100.0}	& \textbf{95.5}	& \textbf{100.0}	& \textbf{100.0}	& - \\
VNS	& \textbf{96.4}	& 96.2	& 85.7	& 92.3	& 60.0	& \textbf{100.0}	& 90.9	& 92.9 & \textbf{100.0} &	- \\ \hline
Npops*	& 28	& 26	& 21	& 13	& 5	& 27	& 22	& 14	& 6 & 	0 \\ \hline

\end{tabular}
}
}
\tabnote{\textsuperscript{*} Number of populations wheren BB algorithm produced a feasible solution, that is, attending constraints of the fixed strata number, and the LH algorithm converged}
\tabnote{\textsuperscript{-} BB algorithm dit not produce feasible solution}

\label{tab:Table10}
\end{table}

\begin{table}[htbp]
\tbl{Percentage of times that algorithm produced the best solution}
{
\footnotesize
\begin{tabular}{l|r} \hline
Algorithm & \% Times best \\  \hline
STRATMH	& 94.2 \\
VNS	& 89.1 \\
KO	& 64.1 \\
BB	& 51.5 \\
LH	& 15.7  \\ \hline
\end{tabular}
}
\label{tab:Table11}
\end{table}

\newpage

Finally, Table \ref{tab:Table12} presents the average and median processing times of the STRATMH, BB, and VNS algorithms. Such analysis was restricted to the three algorithms. Although they are based on stochastic optimization methods (metaheuristics), they demand intensive computation, implying more processing time to produce reasonable quality solutions. The LH and KO algorithms are faster, with execution time of the order of a few seconds; however, in general, they produced solutions of lower quality than those achieved by STRATMH and VNS, as observed in Tables \ref{tab:Table6} to \ref{tab:Table11}.

From the analysis of Table 13, it can be observed that STRATMH is the algorithm that generally requires the least computational time to produce the smallest sample size. The median STRATMH time was less than 4 seconds, regardless of the scenario. Likewise, a value of fewer than 5 seconds in average time was observed, at less than $cv_t$ =3\%.

\begin{table}[htbp]
\tbl{Algorithms STRATMH, BB and VNS - Mean and Median Time (seconds) by $L$ and $cv_t$}
{
\resizebox{\textwidth}{!}{
\begin{tabular}{crrrrrrcrrrrrr} \hline
\multicolumn{1}{l}{} &  && $cv_t$=3.0\%      &     && & \multicolumn{1}{l}{} &     && cvt=5.0\%&     && \\
\multicolumn{1}{l}{} & \multicolumn{1}{c}{Mean}    & \multicolumn{1}{c}{}   & \multicolumn{1}{c}{} & \multicolumn{1}{c}{Median}  & \multicolumn{1}{c}{}   & \multicolumn{1}{c}{}    &       & \multicolumn{1}{c}{Mean}    & \multicolumn{1}{c}{}   & \multicolumn{1}{c}{}  & \multicolumn{1}{c}{Median}  & \multicolumn{1}{c}{}   & \\
$L$     & STRATMH & BB & VNS & STRATMH & BB & VNS & $L$ & STRATMH & BB & VNS & STRATMH & BB & VNS \\ \hline
3     & 3.5 & 31.8    & 48.1  & 0.7 & 22.3    & 15.4     & 3     & 1.5 & 29.3    & 45.1   & 0.6 & 23.3    & 14.9     \\
4     & 8.5 & 40.2    & 69.2  & 1.5 & 24.7    & 21.5     & 4     & 3.3 & 29.9    & 43.1   & 0.9 & 22.5    & 16.6     \\
5     & 13.4& 50.9    & 127.5 & 2.7 & 30.5    & 46.0     & 5     & 4.0 & 50.9    & 105.9  & 1.1 & 27.4    & 58.6     \\
6     & 15.0& 46.2    & 386.5 & 2.0 & 28.2    & 165.5    & 6     & 4.2 & 36.9    & 197.5  & 1.7 & 26.4    & 126.8    \\
7     & 15.7& 52.0    & 2190.3& 3.1 & 28.3    & 760.7    & 7     & 4.9 & 32.9    & 719.8  & 2.0 & 29.0    & 514.5    \\ \hline
\multicolumn{1}{l}{} & \multicolumn{1}{r}{}        & \multicolumn{1}{r}{}   & \multicolumn{1}{r}{cvt=7.5\%} & \multicolumn{1}{r}{}        & \multicolumn{1}{r}{}   & \multicolumn{1}{r}{}    & \multicolumn{1}{r}{} & \multicolumn{1}{r}{}        & \multicolumn{1}{r}{}   & \multicolumn{1}{r}{cvt=10.0\%} & \multicolumn{1}{r}{}        & \multicolumn{1}{r}{}   & \multicolumn{1}{r}{}    \\
\multicolumn{1}{l}{} & \multicolumn{1}{c}{Mean}    & \multicolumn{1}{c}{}   & \multicolumn{1}{c}{} & \multicolumn{1}{c}{Median}  & \multicolumn{1}{c}{}   & \multicolumn{1}{c}{}    &       & \multicolumn{1}{c}{Mean}    & \multicolumn{1}{c}{}   & \multicolumn{1}{c}{}  & \multicolumn{1}{c}{Median}  & \multicolumn{1}{c}{}   & \multicolumn{1}{c}{}    \\
$L$     & STRATMH & BB & VNS & STRATMH & BB & VNS & $L$ & STRATMH & BB & VNS & STRATMH & BB & VNS \\ \hline
3     & 0.8 & 28.8    & 20.3  & 0.6 & 21.2    & 9.5      & 3     & 0.6 & 31.1    & 18.8   & 0.5 & 20.1    & 8.5      \\
4   & 1.3 & 31.6  & 27.4  & 0.9 & 20.2    & 17.3   & 4.0  & 0.9 & 36.4 & 18.3   & 0.8 & 18.9    & 12.5\\
5   & 1.7 & 22.8 & 44.4  & 1.2 & 20.1    & 27.1     & 5   & 1.3 & 20.1 & 27.8 & 1.1 & 20.3  & 18.8     \\
6     & 2.4 & 30.0  & 108.9 & 1.6 & 27.5    & 56.7  & 6  & 2.3 & 29.1 & 35.5 & 1.6 & 26.8    & 32.8     \\
7     & 2.4 & 22.8    & 264.6 & 1.9 & 20.3    & 240.3    & 7     & -       & -  & - & - & -  & - \\ \hline
\end{tabular} 
}}
\tabnote{\textsuperscript{-} BB algorithm did not produce a feasible solution}
\label{tab:Table12}
\end{table}

\section{Conclusions}

This article presented a hybrid method with a new heuristic algorithm to solve the univariate stratification problem. This is an important problem in the field of statistics and has high computational complexity. In order to solve this problem and produce reasonable quality solutions, an algorithm based on BRKGA called STRATMH was proposed, which considers the method proposed by \cite{Brito2015} in the allocation step.

The STRATMH was evaluated and compared with four algorithms from the literature by carrying out experiments considering 31 populations and 20 scenarios, which contemplated different numbers of strata and target coefficients of variation.

 From the analyses, it was possible to observe a superior performance of this algorithm compared to the four algorithms, and this statement is corroborated by the percentage of best solutions (smaller sample sizes) produced by STRATMH in all evaluated scenarios and the three analyzes performed.

In Analysis (I), which involved the KO and VNS algorithms, STRATMH produced 96\% of the best solutions. In Analysis (II), where the same algorithms were considered in Analysis (I) + LH algorithm, STRATMH also produced almost 96\% of the best solutions. Finally, in Analysis (III), where the solutions of the five algorithms were considered, a percentage of 94\% was observed for STRATMH.

Regarding the Analysis of the algorithm's efficiency, translated by the computational time required to produce the solutions, where the comparison between the computational times of STRATMH and the VNS and BB algorithms, also based on metaheuristics, was considered; where STRATMH proved to be the most efficient, taking into account the average and median times.

The results and analyses presented in this work indicate that the STRATMH algorithm is a good alternative for solving the univariate stratification problem to minimize the sample size considering the number of strata and pre-fixed level of precision.

Appendix 2 brings two examples of the application of the function that implements the STRATMH heuristic in two of the 31 populations used in this work.

\section{Appendices}

\appendix

\section{Crossover Procedure}

\begin{figure}[htbp]
    \centering
    \includegraphics[scale=2.0]{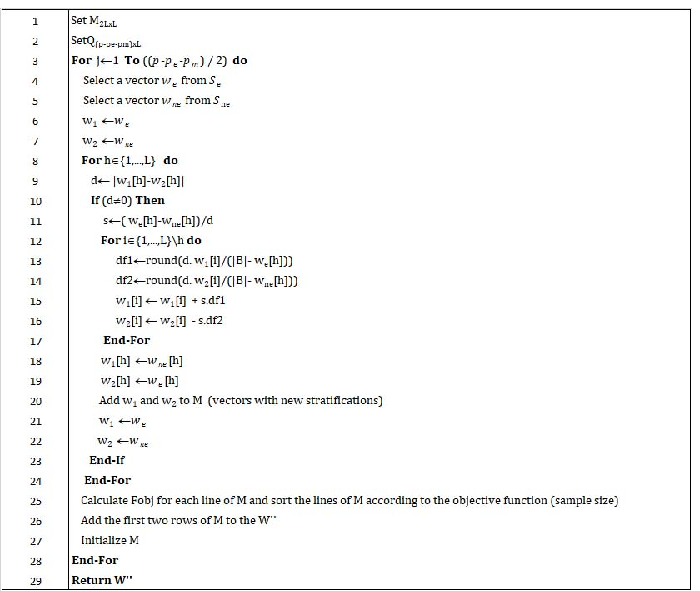}
    \caption{Crossover pseudocode}
    \label{fig:Fig7}
\end{figure}

Figure \ref{fig:Fig7}  shows pseudocode with all the steps of the crossover procedure described in subsection \ref{CMO}. In steps 1 and 2, the matrices where the vectors $w$ corresponding to the new stratifications, produced between lines 3 and 28, are defined. Then, in the main loop from line 3 are selected (lines 4 and 5) vectors from the sets $S_e$ and $S_{ne}$, and  these vectors (in lines 6 and 7) are copied to vectors $w_1$ and $w_2$, which will be changed/updated for each change of values of each of their $L$ positions, from line 8 (intermediate loop). Next, between lines 12 and 19, the values of each of the positions of $w_1$ and $w_2$ are swapped, and the ($L$-1) remaining values are updated, as per lines 13 to 16 (inner loop). Finally, in line 20, the new vectors produced are added to matrix $M$ and then reset with the initial values of vectors $w_e$ and $w_{ne}$. At the end of the intermediate loop, after line 24, the values of $N_h$, $S_{hx}^2$, and the sample sizes corresponding to the 2*$L$ stratifications associated with $M$ are calculated.

Finally, in line 26, the two best stratifications, with the smallest associated sample sizes, are added to the matrix $Q$, which contains all the ($p-p_e-p_m$) solutions obtained from the crossover.

\section{stratMH Function}

The STRATMH algorithm was implemented in R language and is available in the stratMH function (see Table \ref{tab:Table3}), which makes use of the parallel and MultAlloc packages (with the implementation of the allocation method proposed in \cite{Brito2015}). Tables \ref{tab:p1} and \ref{tab:p2} bring, respectively, information about the parameters used as input to the function and the results produced from its application. Finally, Table \ref{tab:p3} brings two examples of the application of this function.

\begin{table}[htbp]
\tbl{stratMH Function - Input Parameters}
{\begin{tabular}{l|l} \hline
Parameters &	Description \\ \hline
X	& Vector with $N$ observations of the stratification variable \\
L	& Number of strata (default value  = 3) \\
cvt	& Target cv ( default value = 0.1 – 10\%) \\
p	& Number of solutions ($w$ vectors) per generation (default value = 50) \\
pe	& Percentage of chromosomes in Se (default value = 0.3) \\
pm	& Percentage of chromosomes produced in the mutation (default value  = 0.3) \\
maxgen	& Maximum Heuristic Generations (default value = 50) \\
npar	& Performs the number of executions of the algorithm equal   \\
  &   to npar (default  = 1), npar $>$ 1  parallelism \\ \hline
\end{tabular}
} 
\label{tab:p1}
\end{table}

\begin{table}[htbp]
\tbl{Results Produced by the stratMH Function}
{\begin{tabular}{l|l} \hline
Values	& Description \\ \hline
xbest	& Vector with $L$ values of $w$ that satisfy equations (\ref{eq:n17}) e (\ref{eq:n18}) \\
bk	& Vector with  $L-1$  cutoff $b_1$, $b_2$, \dots, $b_{L-1}$ \\
Nh	& Vector with number of observations of $X$ in each stratum  $L$ stratums \\
nh	& Vector with sample size allocated to each stratum \\
Sh2	& Vector with population variance concerning to variable $X$ in each of $L$ strata \\
fobj	& Total sample size ($n$) \\
cv	& Coefficient of variation obtained from the stratification produced \\
ta	& Runtime in seconds \\
ibest	& The generation where was produced the best solution  (smallest sample size) \\ \hline
\end{tabular}
}
\label{tab:p2}
\end{table}

\begin{table}[htbp]
\tbl{Examples of the application of the stratMH Function}
{\begin{tabular}{l} \hline
Example 1: Population - Debtors \\ \hline
X=scan(“Debtors.txt”,dec=”.”) \\
s=stratMH(X,L=4,cvt=0.05) \#target cv of 5.0\%  \hspace{3cm}  \\
s \\
\$xbest \\
{[1]} 337 459 261  72 \\
\$bk \\
{[1]}  437 1705 6029 \\
\$Nh \\
{[1]} 2079  910  308   72 \\
\$nh \\
{[1]} 13 16 20 20 \\
\$Sh2 \\
{[1]}    12219.96   104620.51  1346923.45 24705371.94 \\
\$fobj \\
{[1]} 69 \\
\$cv \\
{[1]} 0.04999865 \\
\$ta \\
elapsed \\
   1.62 \\
\$ibest \\
{[1]} 21 \\ \hline

Example 2: Population - BeefFarms \\ \hline
X=scan("BeefFarms.txt",dec=".") \\
s=stratMH(X,L=3,cvt=0.01,npar=3) \#target cv of 1.0\% \\
s \\
\$xbest \\
{[1]} 163  87 103 \\
\$bk \\
{[1]} 323 884 \\
\$Nh \\
{[1]} 228  97 105 \\
\$nh \\
{[1]}  16  16 105 \\
\$Sh2 \\
{[1]}     5208.898    27054.350 22737213.792 \\
\$fobj \\
{[1]} 137 \\
\$cv \\
{[1]} 0.009932989 \\
\$ta \\
elapsed\\
   4.63 \\
\$ibest \\
{[1]} 4 \\ \hline

\end{tabular}}
\label{tab:p3}
\end{table}

\end{document}